\pdfoutput=1
\documentclass[preprint,preprintnumbers,amsmath,amssymb,floatfix,endfloats*]{revtex4}

\usepackage{graphicx}
\usepackage{dcolumn}
\usepackage{bm}
\usepackage{amsfonts}

\DeclareMathAlphabet{\mathsfsl}{OT1}{cmr}{bx}{it}
\begin{document}
\title{Molecular dynamics simulation study of a polymer droplet transport over an array of spherical nanoparticles}
\author{Anish Thomas and Nikolai V. Priezjev}
\affiliation{Department of Mechanical and Materials Engineering,
Wright State University, Dayton, Ohio 45435}
\date{\today}
%
\begin{abstract}

The dynamic behavior of a partially wetting polymer droplet driven
over a nanostructured interface is studied using molecular dynamics
simulations.   We consider the bead-spring model to represent a
polymeric liquid that partially wets a rough surface composed of a
periodic array of spherical particles.  It is shown that at
sufficiently small values of the external force, the droplet remains
pinned at the particles' surface, while above the threshold, its
motion consists of alternating periods of pinning and rapid
displacements between neighboring particles. The latter process
involves large periodic variation of the advancing and receding
contact angles due to attachment and detachment of the contact line.
Finally, upon increasing external force, the droplet center of mass
is displaced steadily but the oscillation amplitude of the receding
contact angle as well as the maximum contact angle hysteresis remain
relatively unaffected.

\vskip 0.5in


\end{abstract}

\maketitle

\section{Introduction}

The development of nanofabrication techniques is important for
manufacturing the so-called \textit{superhydrophobic} surfaces that
possess liquid-repellent and self-cleaning properties~\cite{Sam19}.
The main design principle behind such surfaces is to introduce
small-scale surface asperities that can keep the liquid interface
suspended above the substrate and allow for the formation of trapped
air pockets, thus reducing the liquid-solid contact
area~\cite{Jeevahan18,Phuong19}. This situation is commonly referred
to as the Cassie-Baxter state~\cite{CB44}, whereas the case of a
full penetration of a liquid into a rough surface is called the
Wenzel state~\cite{Wenzel36}. As a result, liquids in contact with
superhydrophobic surfaces typically exhibit large contact angles,
relatively small contact angle hysteresis, and reduced hydrodynamic
friction.  The enhanced slip properties also offer a more precise
control of liquid flow in microfluidic and nanofluidic
channels~\cite{Vinograd12,Neto14}. In particular, it was recently
shown that the effective slip length for liquid flows over
substrates with mixed boundary conditions can be accurately computed
using either continuum or atomistic simulations, provided that the
length scales of surface textures are larger than the molecular
size~\cite{Priezjev05,Priezjev11,Bao18suphyd,Bao20SH}.

\vskip 0.05in

In recent years, the wetting and flow properties of liquid droplets
over nanotextured surfaces were examined using molecular dynamics
simulations~\cite{Servantie08,Koishi09,Yong09,ZhaoSci13,Singh14,
Gendt14,Tang14,Pei15,ZhaoNano15,Wang15,Das16,Ma16,Tsao16,JYan17,Yen17,Yu18,Hendy18,
Prie20Geng}.   In the absence of external forces, the stability of
the Wenzel or Cassie-Baxter states depends on the height of surface
pillars, their surface fraction, and the intrinsic contact
angle~\cite{Koishi09,Singh14,Tang14,Wang15}.  In turn, the
Cassie-to-Wenzel wetting transition at pillar-arrayed surfaces and
subsequent spreading of the liquid can be facilitated by applying a
sufficiently large electric field~\cite{ZhaoNano15,Das16}.
Furthermore, under an external force parallel to nano-pillared
superhydrophobic surface, the advancing and receding contact lines
of a water droplet are displaced in discrete steps, and the
partially wetting droplet effectively rolls above the pillar
tips~\cite{Yu18,Hendy18}. More recently, it was shown that directed
droplet motion along a hydrophobic stripe can be generated by
periodic vibration of a substrate with asymmetric corrugations, and
the droplet velocity follows a power-law dependence on the vibration
period~\cite{Prie20Geng}.  However, despite extensive efforts, the
exact mechanisms of the contact line displacement, variation of the
advancing and receding contact angles, and neck formation during
pinning of a partially wetting droplet at nanostructured surfaces
remain not fully understood.

\vskip 0.05in

In this paper, molecular dynamics simulations are carried out to
investigate the effects of an external force and wettability on the
displacement of a polymer droplet over a smooth substrate covered by
periodically arranged spherical nanoparticles. It will be shown that
below the threshold force, the partially wetting droplet remains
pinned at the particles' surface, while at larger values of the
external force, the droplet motion becomes intermittent and involves
rapid displacements separated by periods of pinning. The process of
attachment and detachment from the particle surface is associated
with large variation of the advancing and receding contact angles.
Interestingly, the maximum value of the contact angle hysteresis
remains rather insensitive to the external force.

\vskip 0.05in

The rest of the paper is structured as follows. The details of the
molecular dynamics simulation model and parameter values are given
in the next section.  The description of the fitting procedure for
the droplet shape, examples of the droplet profiles at structured
interfaces, and the time dependence of the droplet displacement as
well as the variation of the advancing and receding contact angles
are presented in Sec.\,\ref{sec:Results}.  The last section contains
a brief summary of the results.

\section{Molecular Dynamics Simulations}
\label{sec:MD_Model}


In our study, a liquid droplet consists of $880\,000$ atoms
interacting via the Lennard-Jones (LJ) potential. More specifically,
any two atoms interact via the truncated LJ potential:
\begin{equation}
V_{LJ}(r)=4\,\varepsilon\,\Big[\Big(\frac{\sigma}{r}\Big)^{12}\!-\Big(\frac{\sigma}{r}\Big)^{6}\,\Big],
\label{Eq:LJ}
\end{equation}
where the parameters $\varepsilon$ and $\sigma$ describe the energy
and length scales of the liquid phase.   The interaction between
solid and liquid atoms is also described by the LJ potential with
the parameters $\varepsilon_{\rm sp}$ (measured in units of
$\varepsilon$) and $\sigma_{\rm sp}=\sigma$. To speed up
simulations, the cutoff radius $r_{c}=2.5\,\sigma$ is used for all
types of interactions. The equations of motion were solved using the
velocity Verlet integration algorithm~\cite{Allen87} with the time
step $\triangle t_{MD}=0.005\,\tau$, where
$\tau=\sigma\sqrt{m/\varepsilon}$ is the LJ time.  The molecular
dynamics simulations were carried out using the LAMMPS parallel
code~\cite{Lammps}.

\vskip 0.05in


In addition to the LJ interaction, liquid monomers are connected to
form short chains (10 monomers per chain) via the FENE (finitely
extensible nonlinear elastic) potential:
\begin{equation}
V_{FENE}(r)=-\frac{k}{2}\,r_{\!o}^2\ln[1-r^2/r_{\!o}^2],
\label{Eq:FENE}
\end{equation}
with the parametrization $k\,{=}\,30\,\varepsilon\sigma^{-2}$ and
$r_{\!o}\,{=}\,1.5\,\sigma$~\cite{Kremer90}.   Thus, the neighboring
atoms within a chain interact via the LJ and FENE potentials, which
effectively yields a harmonic potential that prevents polymer chains
from unphysical crossing each other~\cite{Kremer90}. The temperature
of the liquid phase, $T=1.0\,\varepsilon/k_B$, was regulated via the
Nos\'{e}-Hoover thermostat applied along the $\hat{y}$ direction
(perpendicular to the droplet direction of motion). Here, $k_B$ is
the Boltzmann constant.

\vskip 0.05in

\begin{figure}[t]
\includegraphics[width=12.cm,angle=0]{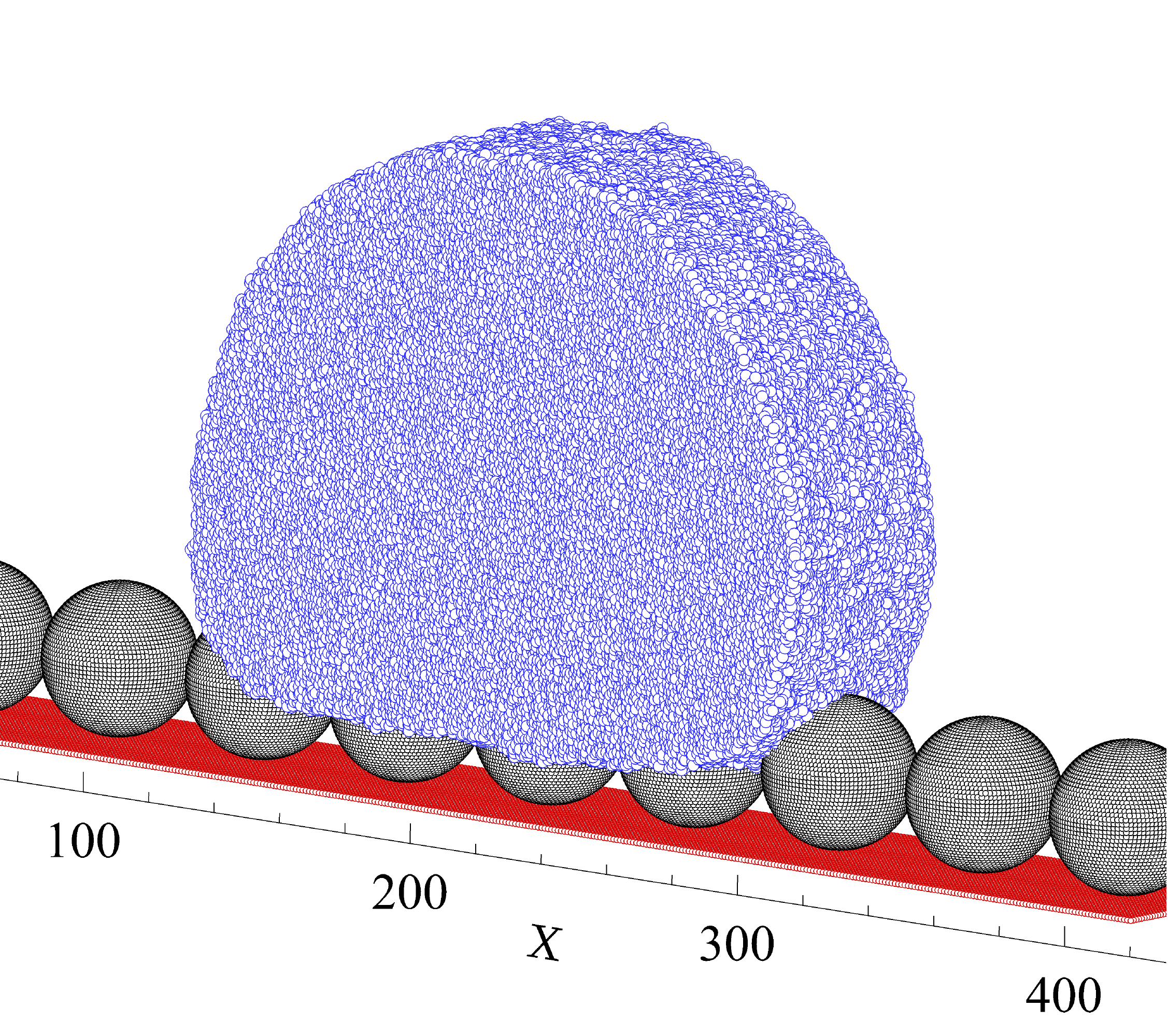}
\caption{(Color online) An instantaneous snapshot of the liquid
droplet that consists of $88\,000$ polymer chains (10 monomers per
chain) and resides on the array of spherical particles rigidly
attached to the solid substrate. The atoms are not shown to scale.
The interaction energy between liquid monomers and particle atoms is
$\varepsilon_{\rm sp}\!=\!0.6\,\varepsilon$. The external force
$f_x=0.00004\,\varepsilon/\sigma$ is applied on each monomer. }
\label{fig:snapshot_droplet}
\end{figure}

The polymer droplet is initially placed in contact with a composite
substrate, which consists of an array of 10 spherical particles
rigidly attached to a stationary solid plane, as shown in
Fig.\,\ref{fig:snapshot_droplet}. The solid plane consists of
$22\,000$ atoms arranged on a square lattice within the $xy$ plane
with lateral dimensions $440\,\sigma \times 50\,\sigma$. The atoms
in the solid plane do not interact with each other and they are
rigidly fixed to the lattice sites.   Next, each spherical particle
is composed of $4000$ atoms uniformly distributed on a surface of a
sphere with the radius $R=17.8\,\sigma$.   The areal density of
atoms in the solid plane and spherical particles is
$1.0\,\sigma^{-2}$.  In our setup, the atoms on the surface of
spherical particles do not interact with each other but the
interaction energy between liquid monomers and particle atoms is
controlled by the parameter $\varepsilon_{\rm sp}$ (measured in
units of $\varepsilon$).

\vskip 0.05in


The preparation procedure involved a thermal equilibration of the
polymeric liquid starting from the crystal phase. Periodic boundary
conditions were imposed along the $\hat{x}$ and $\hat{y}$
directions, while the system is open along the $\hat{z}$ direction
(perpendicular to the solid plane).  After the initial
equilibration, the liquid droplet was gradually displaced towards
the composite substrate and further allowed to equilibrate in a
partially wetting state at a given value of $\varepsilon_{\rm sp}$.
The droplet motion along the $\hat{x}$ direction was induced by a
constant force, $f_x$, applied on each monomer. It should be noted
that the force of gravity was not included in our study.

\section{Results}
\label{sec:Results}

\begin{figure}[t]
\includegraphics[width=14.cm,angle=0]{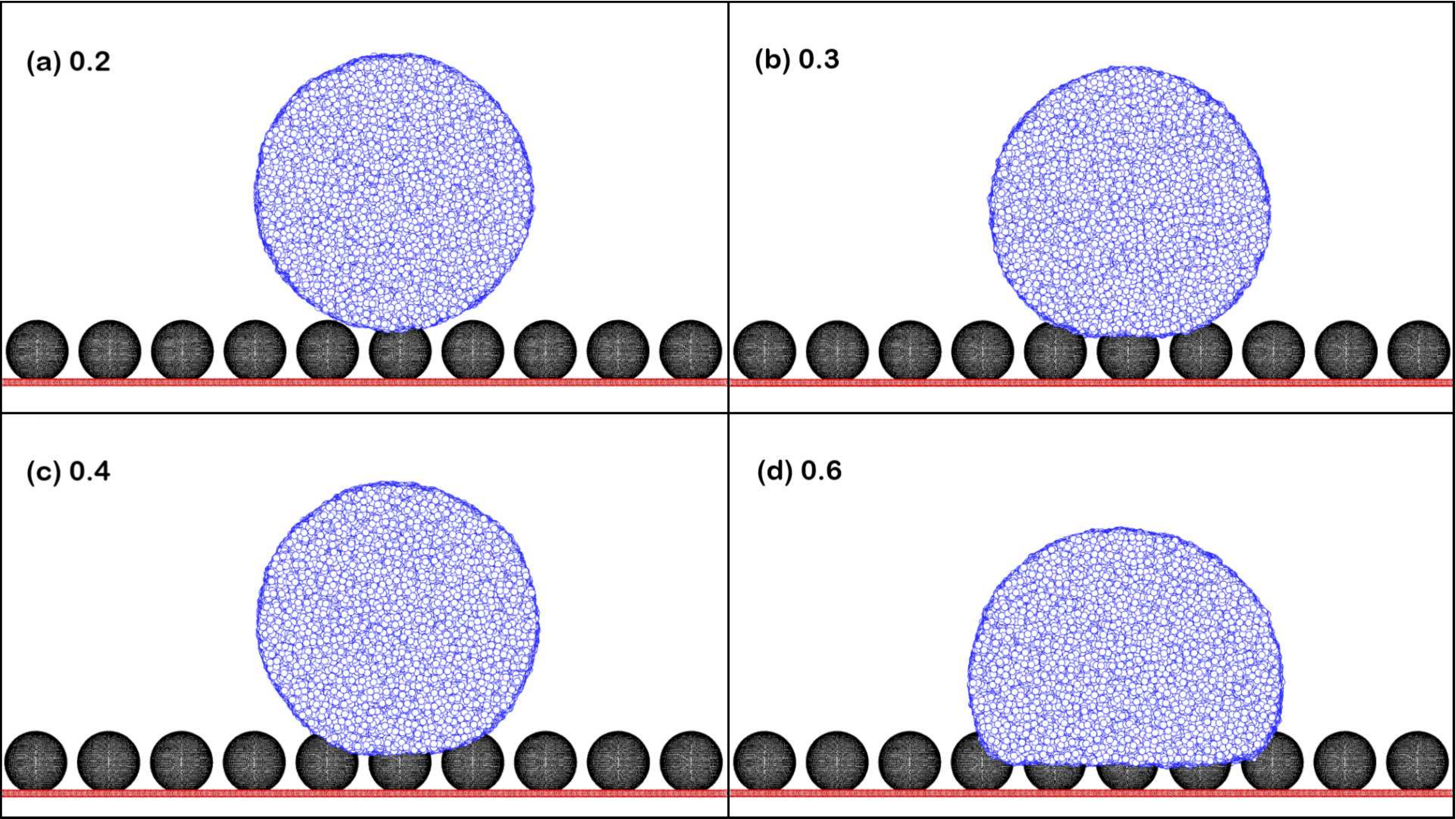}
\caption{(Color online) The side view of polymer droplets in contact
with solid particles for the surface energies (a) $\varepsilon_{\rm
sp}\!=\!0.2\,\varepsilon$, (b) $\varepsilon_{\rm
sp}\!=\!0.3\,\varepsilon$, (c) $\varepsilon_{\rm
sp}\!=\!0.4\,\varepsilon$, and (d) $\varepsilon_{\rm
sp}\!=\!0.6\,\varepsilon$. }
\label{fig:eq_shapes}
\end{figure}

It is well realized that the wetting properties of structured
interfaces depend sensitively on a number of factors, \textit{i.e.},
the areal density and shape of the surface texture, the surface
energy, and the surface tension at the liquid-vapor
interface~\cite{McKinley07}.  In the present study, the surface
roughness is introduced via an array of spherical particles, and the
liquid is confined into a narrow slab with periodic boundary
conditions to access larger droplet sizes. In turn, the liquid
consists of short, flexible polymer chains in order to enhance
surface tension and, thus, reduce evaporation from the liquid
interface. The examples of droplet snapshots in contact with the
periodic array of particles are displayed in
Fig.\,\ref{fig:eq_shapes} for selected values of the interaction
energy between the liquid monomers and particle atoms. It can be
seen that the droplet shape varies from nearly circular for
$\varepsilon_{\rm sp}\!=\!0.2\,\varepsilon$ to semicircular for
$\varepsilon_{\rm sp}\!=\!0.6\,\varepsilon$. In the latter case, the
linear size of the droplet along the $\hat{x}$ direction is about
$180\,\sigma$. Taking $\sigma=0.5\,\text{nm}$, the typical size of
the droplet is about $90\,\text{nm}$ and the sphere radius is
$\approx\!9\,\text{nm}$.

\vskip 0.05in

\begin{figure}[t]
\includegraphics[width=12.cm,angle=0]{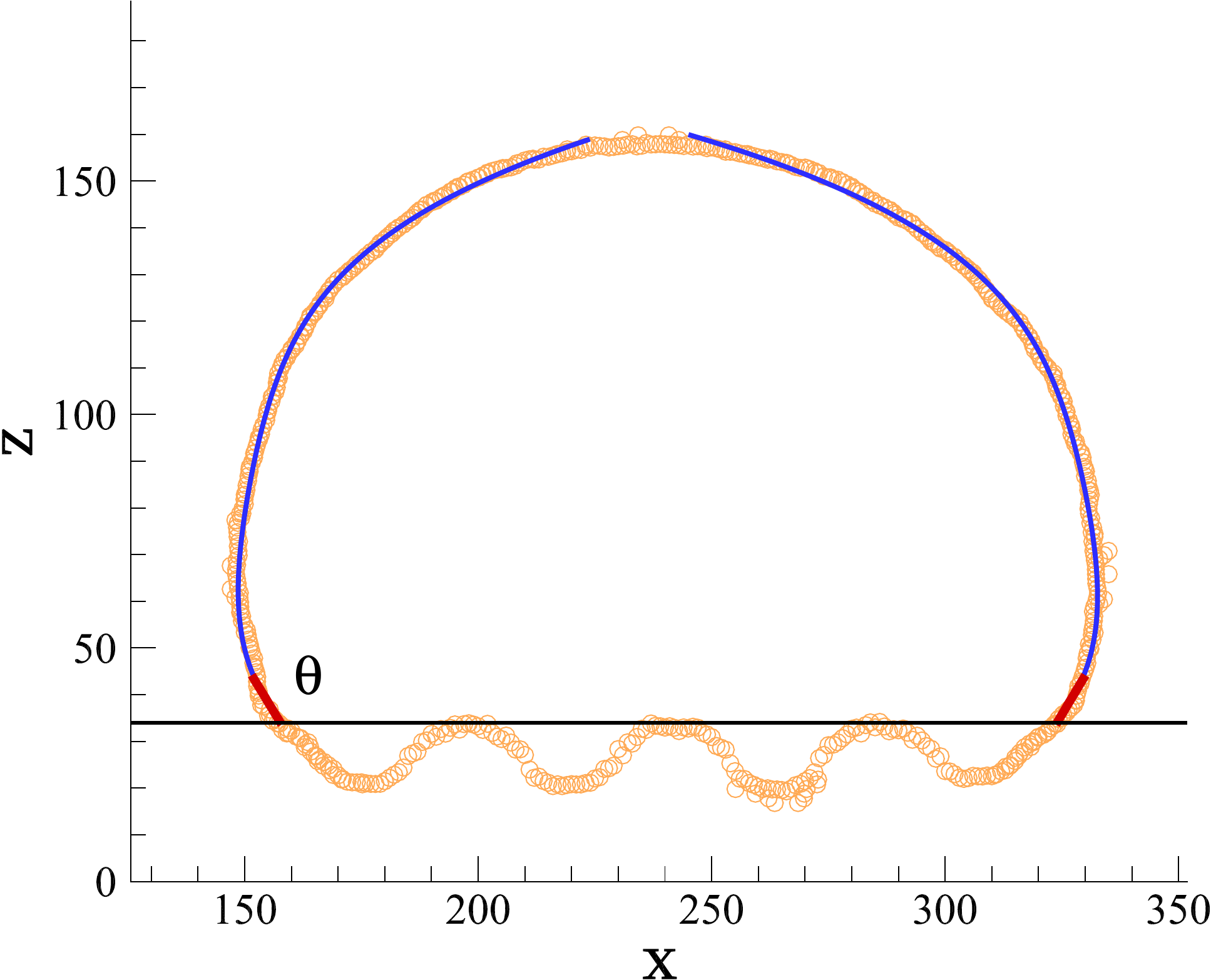}
\caption{(Color online) An example of the droplet interface (orange
circles) computed from the spatial variation of the local density
profiles (see text for details) for $\varepsilon_{\rm
sp}\!=\!0.6\,\varepsilon$ and $f_x=0$. The horizontal black line
denotes the reference plane above the tips of spherical particles.
The blue curves indicate the 4-th order polynomial fit, and the red
lines denote the local tangents to the polymer interface used to
compute the apparent contact angles. }
\label{fig:fitting}
\end{figure}

We next describe the fitting procedure used to obtain the droplet
profiles and apparent contact angles.  The difficulty in computing
the position of the liquid interface from instantaneous
configuration of monomers arises due to thermal fluctuations that
result in a finite width of the interface. Therefore, the
computational domain was divided into narrow slices (with thickness
of $1.0\,\sigma$) parallel and perpendicular to the solid wall. In
each slice, the local density of liquid monomers was computed and
the location of the droplet interface was determined at half the
density of liquid monomers in the center of the droplet. An example
of the two-dimensional droplet interface profile is presented in
Fig.\,\ref{fig:fitting}. In general, the definition of the apparent
contact angle at the atomic scale can be ambiguous. In our setup, we
choose a plane located on the distance $2R$ from the solid wall as a
reference for computing the apparent contact angles (see the
horizontal black line in Fig.\,\ref{fig:fitting}). Next, the front
and back portions of the interface were obtained via the 4-th order
polynomial fit, and the tangent at the intersection with reference
plane was computed using the first 10 data points. The accuracy of
this procedure was tested by trial and error for various shapes of
the droplet during its motion over an array of particles. In what
follows, we define the advancing, $\theta_a$, and receding,
$\theta_r$, contact angles at the front and back of the droplet with
respect to the direction of the external force.

\vskip 0.05in

\begin{figure}[t]
\includegraphics[width=12.cm,angle=0]{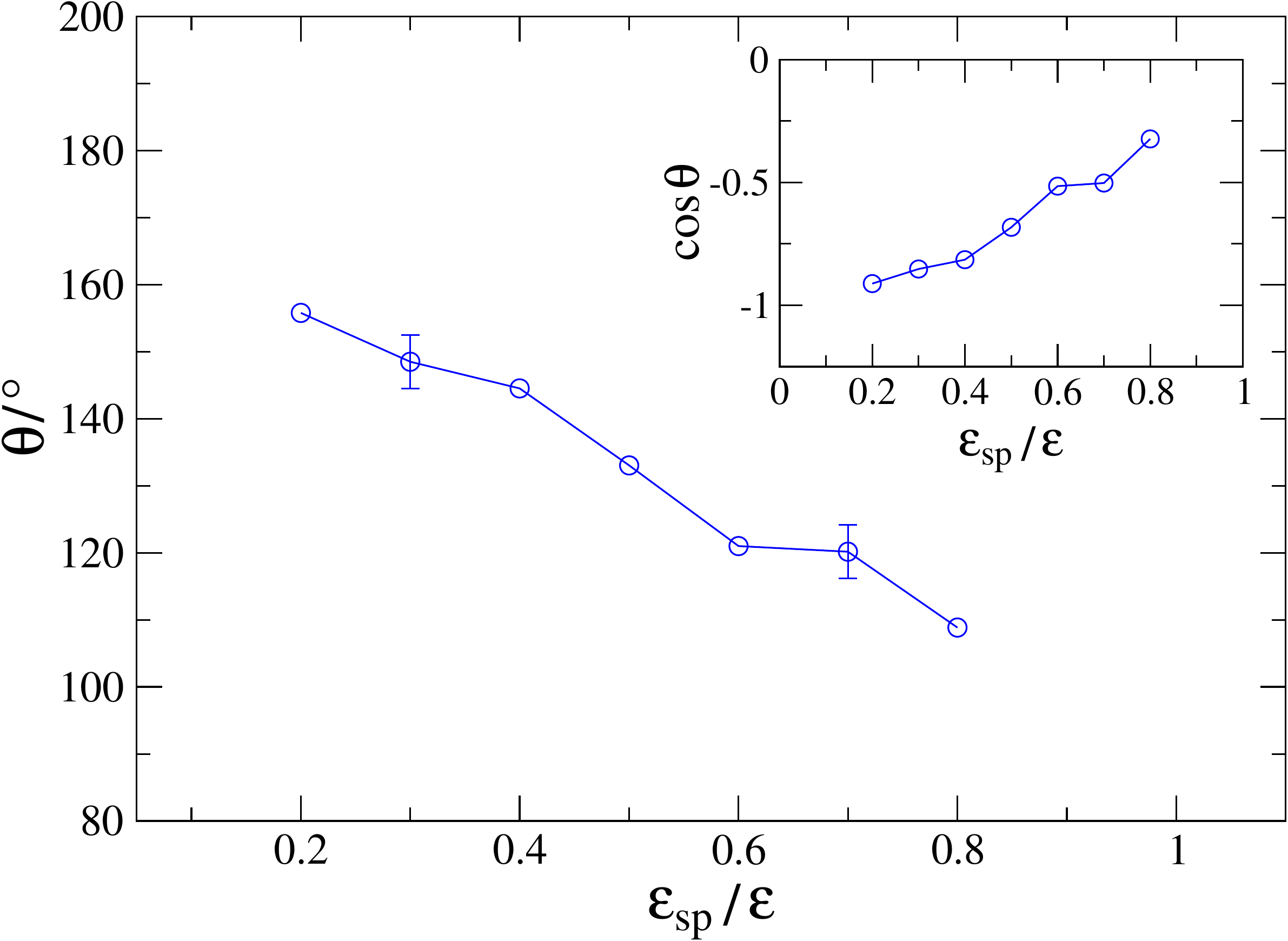}
\caption{(Color online) The apparent contact angle for the droplet
suspended on an array of spherical particles as a function of the
interaction energy between sphere atoms and liquid monomers. The
same data are replotted in the inset as
$\text{cos}\,\theta(\varepsilon_{\rm sp})$.}
\label{fig:contact_angle_eps}
\end{figure}

The dependence of the apparent (static) contact angle as a function
of the interaction energy between sphere atoms and droplet monomers,
$\varepsilon_{\rm sp}/\varepsilon$, is shown in
Fig.\,\ref{fig:contact_angle_eps} in the absence of the external
force. As is evident, the contact angle decreases monotonically with
increasing surface energy; however, the droplet remains in a
partially wetting state for $\varepsilon_{\rm
sp}\leqslant0.8\,\varepsilon$. Note that these results are obtained
for a specific areal fraction, \textit{i.e.}, $\phi_{S} = 10\,\pi
R^2 / A \approx 0.48$, where $A=50\times440\,\sigma^2$ is the area
of the solid plane, and $R=18.3\,\sigma$ is the radius of a sphere
with the excluded volume of atoms of size $\sigma$.   In the recent
study, it was shown that the critical pressure, below which a
polymer film remains suspended on an array of spherical particles
with a similar areal fraction, can be equally well estimated by MD
simulations and numerical minimization of the interfacial
energy~\cite{Bishal18}. We comment also that the local contact angle
of a polymer (10-mers) droplet residing on a flat solid plane with
the density of $1.0\,\sigma^{-2}$ was evaluated previously as a
function of the surface energy~\cite{Bishal18}.

\vskip 0.05in

\begin{figure}[t]
\includegraphics[width=12.cm,angle=0]{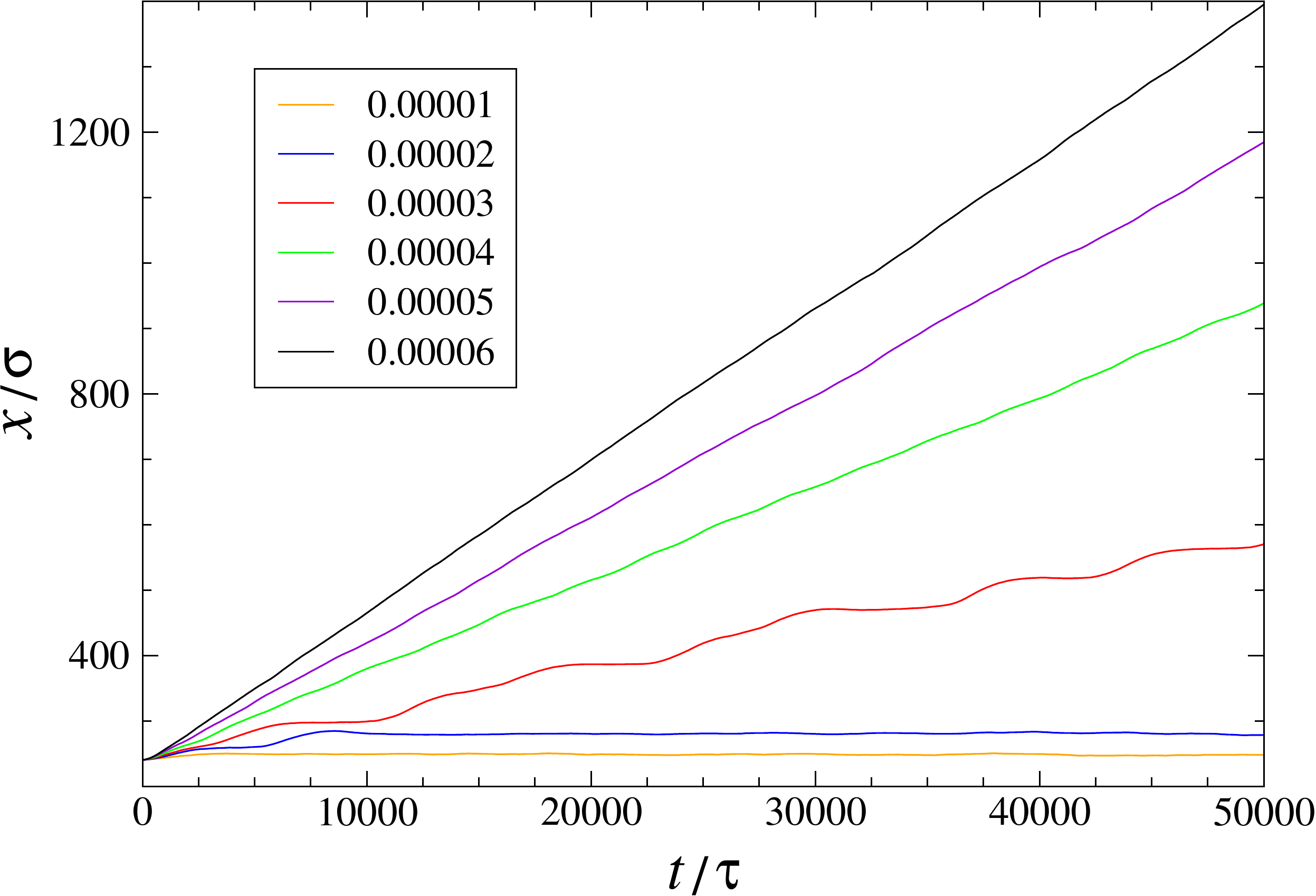}
\caption{(Color online) The time dependence of the droplet center of
mass for the indicated values of the external force, $f_x$ (in units
$\varepsilon/\sigma$). The interaction energy between sphere atoms
and liquid monomers is $\varepsilon_{\rm sp}\!=\!0.6\,\varepsilon$.
}
\label{fig:x_CM_time_fx}
\end{figure}

We next study the droplet displacement under external force applied
to each monomer along the $\hat{x}$ direction.
Fig.\,\ref{fig:x_CM_time_fx} shows the time dependence of the
droplet center of mass for the indicated values of the external
force.  In all cases, the droplet is initially at equilibrium with
the structured interface, and the external force, $f_x$, is suddenly
applied to each monomer at time $t=0$.  It can be clearly observed
that at small values of the force,
$f_x\leqslant0.00002\,\varepsilon/\sigma$, the droplet is initially
displaced on the distance between neighboring spheres but it remains
pinned during the time interval $50\,000\,\tau$.  At the larger
value of the external force, $f_x=0.00003\,\varepsilon/\sigma$, the
droplet becomes temporarily pinned at the particle surface and then
rapidly moves along the array of particles. Upon further increasing
force, the motion becomes more steady although small fluctuations
remain noticeable in the cases $f_x=0.00004\,\varepsilon/\sigma$ and
$0.00005\,\varepsilon/\sigma$. Finally, when
$f_x=0.00006\,\varepsilon/\sigma$, the droplet is displaced linearly
on the distance of about 6 times its size during $50\,000\,\tau$.

\vskip 0.05in

\begin{figure}[t]
\includegraphics[width=12.cm,angle=0]{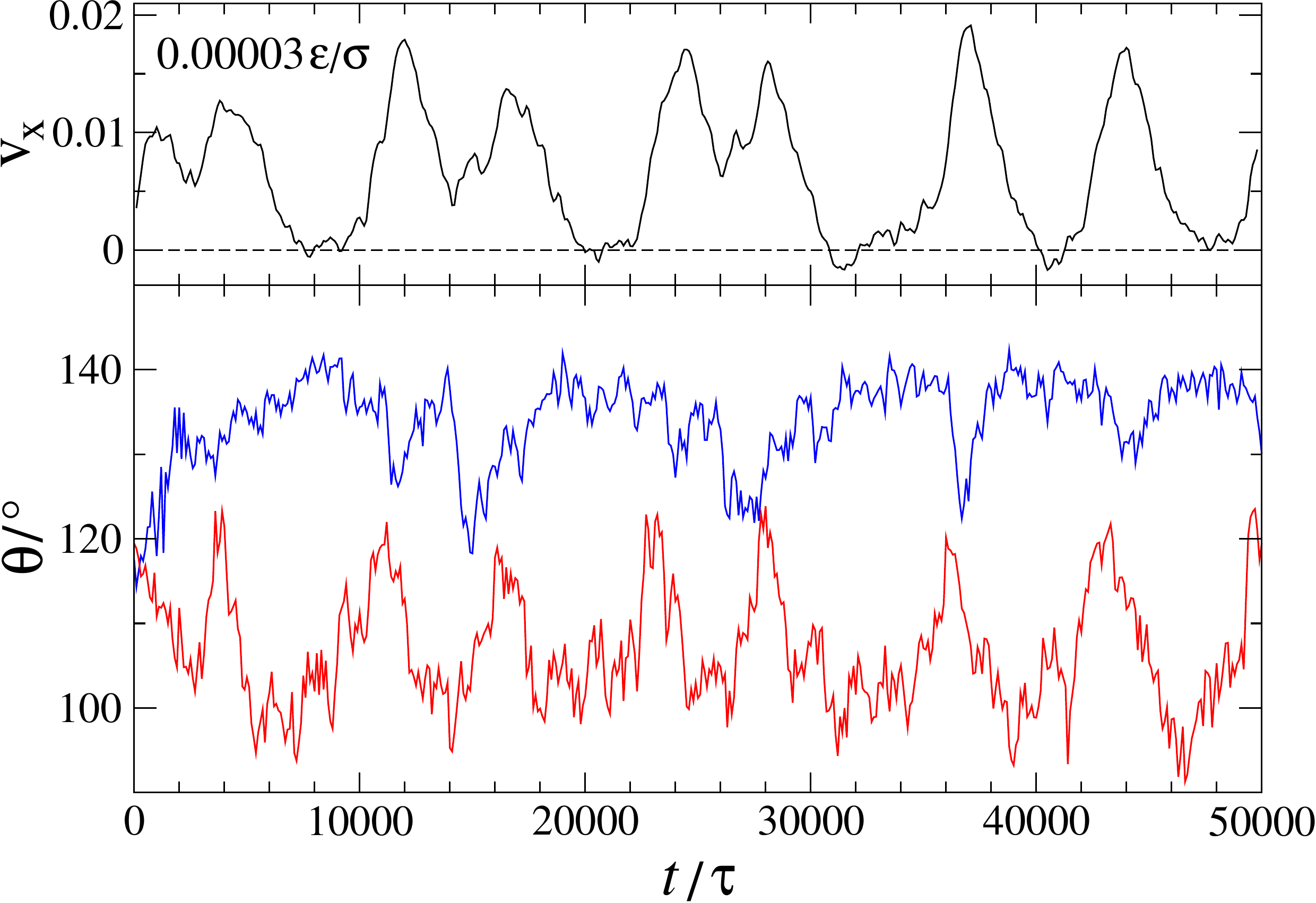}
\caption{(Color online) The velocity of the center of mass (in units
$\sigma/\tau$) as a function of time when
$f_x=0.00003\,\varepsilon/\sigma$ (the upper panel). The horizontal
dashed line indicates $v_x=0$. The advancing (blue curve) and
receding (red curve) contact angles during $50\,000\,\tau$ (the
lower panel). The surface energy is $\varepsilon_{\rm
sp}\!=\!0.6\,\varepsilon$. }
\label{fig:th_vx_time_eps06_fx03}
\end{figure}
\begin{figure}[t]
\includegraphics[width=12.cm,angle=0]{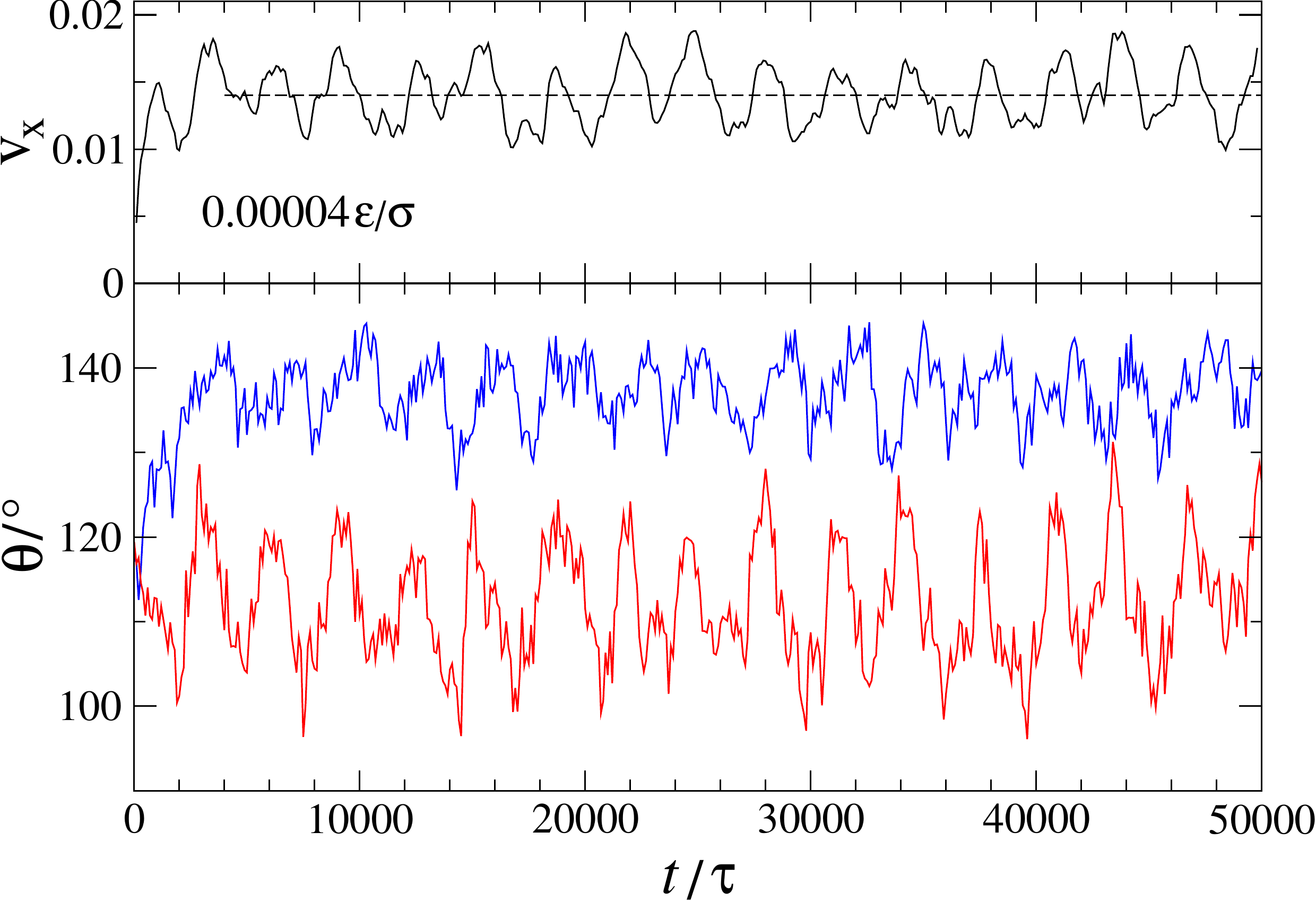}
\caption{(Color online) The center of mass velocity (in units
$\sigma/\tau$) as a function of time when
$f_x=0.00004\,\varepsilon/\sigma$ (the upper panel). The dashed line
denotes the average velocity $v_x=0.014\,\sigma/\tau$. The
interaction energy between sphere atoms and droplet monomers is
$\varepsilon_{\rm sp}\!=\!0.6\,\varepsilon$. The time dependence of
the advancing (blue curve) and receding (red curve) contact angles
for $f_x=0.00004\,\varepsilon/\sigma$ (the lower panel). }
\label{fig:th_vx_time_eps06_fx04}
\end{figure}
\begin{figure}[t]
\includegraphics[width=12.cm,angle=0]{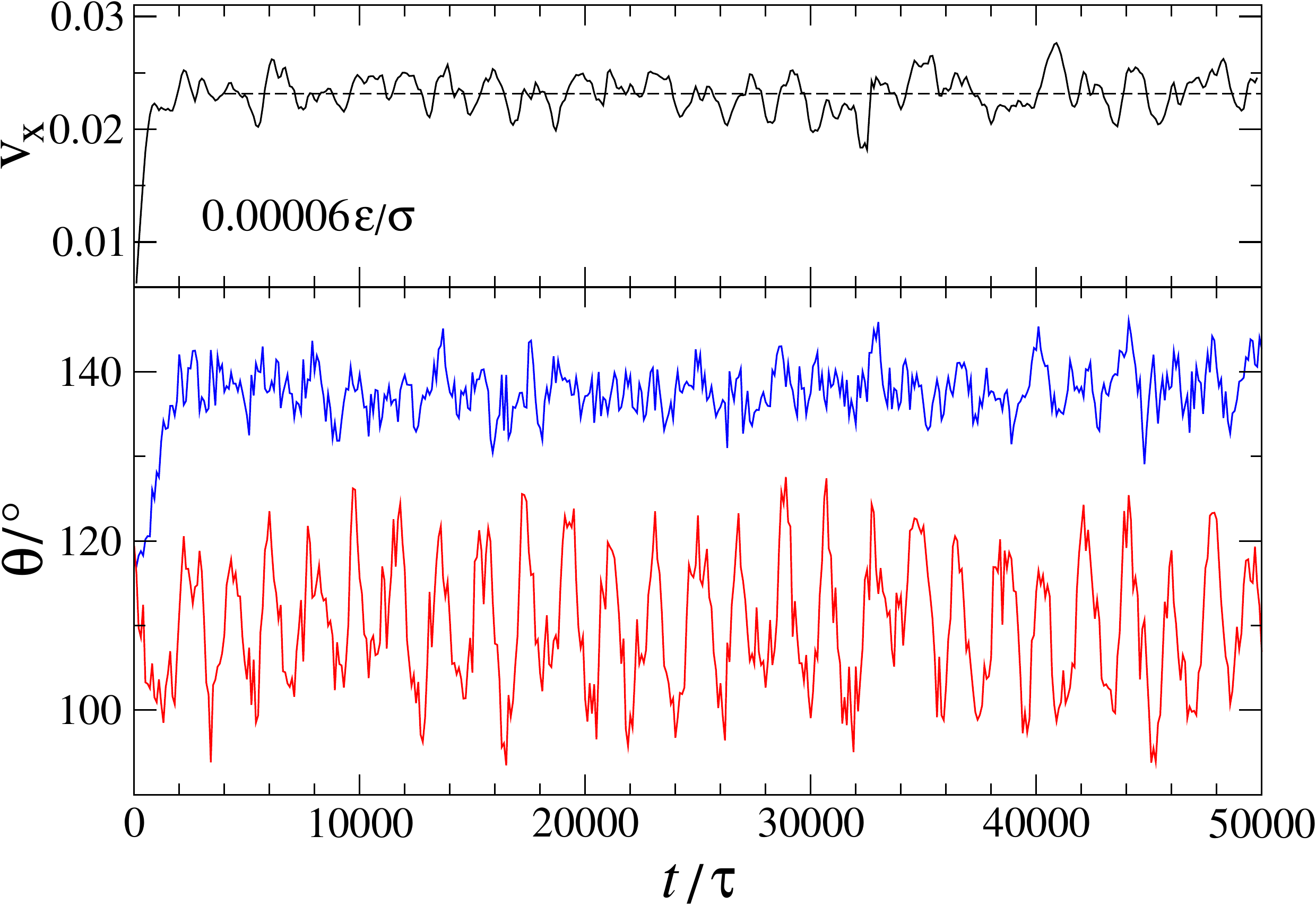}
\caption{(Color online) The upper panel shows the velocity of the
center of mass (in units $\sigma/\tau$) versus time for
$f_x=0.00006\,\varepsilon/\sigma$. The average velocity
$v_x=0.023\,\sigma/\tau$ is indicated by the dashed line. The
surface energy is $\varepsilon_{\rm sp}\!=\!0.6\,\varepsilon$. The
lower panel displays the advancing (blue curve) and receding (red
curve) contact angles as a function of time. }
\label{fig:th_vx_time_eps06_fx06}
\end{figure}

An additional insight into the droplet dynamics can be gained by
analyzing the time dependence of the advancing and receding contact
angles. In
Figures\,\ref{fig:th_vx_time_eps06_fx03}--\ref{fig:th_vx_time_eps06_fx06},
the variation of the advancing and receding contact angles as well
as the velocity of the center of mass are reported for selected
values of the external force. In the case
$f_x=0.00003\,\varepsilon/\sigma$ shown in
Fig.\,\ref{fig:th_vx_time_eps06_fx03}, it can be observed that
droplet dynamics consists of alternating periods of rapid motion and
temporal pinning at the surface of spheres. In general, the maximum
of the center of mass velocity correlates well with the abrupt
changes in the contact angles. Namely, the advancing contact angle
is reduced, while the receding angle increases during transient
motion along the array of spherical particles. Interestingly, one
can also notice in Fig.\,\ref{fig:th_vx_time_eps06_fx03} that the
center of mass velocity occasionally becomes negative between
successive displacements, indicating a slight oscillatory movement
due to surface tension forces.  In addition, the contact angle
hysteresis varies from $\theta_a-\theta_r\approx45^{\circ}$ during
periods of pinning to nearly zero during rapid displacements (see
Fig.\,\ref{fig:th_vx_time_eps06_fx03}).

\vskip 0.05in


Figure\,\ref{fig:th_vx_time_eps06_fx04} shows the droplet velocity
and the contact angles for a larger value of the external force
$f_x=0.00004\,\varepsilon/\sigma$. It can be seen that after the
first jump, the droplet is steadily propelled with the average
velocity $v_x=0.014\,\sigma/\tau$ along the array of particles, and
the amplitude of oscillations of the contact angles is reduced, so
that the advancing contact angle is greater than the receding angle.
The period of the velocity and contact angles variation is
determined by the center of mass velocity and the distance between
neighboring spheres. Note also that the maximum value of the contact
angle hysteresis is nearly the same as in the case
$f_x=0.00003\,\varepsilon/\sigma$, \textit{i.e.,}
$\theta_a-\theta_r\approx45^{\circ}$.

\vskip 0.05in


Furthermore, when the external force is
$f_x=0.00006\,\varepsilon/\sigma$, the average droplet velocity
becomes larger, $v_x=0.023\,\sigma/\tau$, and the fluctuations of
$v_x$ and $\theta_a$ are significantly reduced, as shown in
Fig.\,\ref{fig:th_vx_time_eps06_fx06}. Notice, however, that the
oscillation amplitude of the receding contact angle remains
relatively large and it is comparable to the cases
$f_x=0.00003\,\varepsilon/\sigma$ and $0.00004\,\varepsilon/\sigma$.
In general, the maximum contact angle hysteresis remains rather
insensitive to the value of the external force, as shown in
Figs.\,\ref{fig:th_vx_time_eps06_fx03}--\ref{fig:th_vx_time_eps06_fx06}.
In the previous MD study, the values of surface tension
$\gamma=0.85\,\varepsilon/\sigma^2$ and viscosity
$\eta=11.1\,m/\tau\sigma$ were computed for a thin polymer film of
bead-spring chains (10-mers) at $T=1.0\,\varepsilon/k_B$ and zero
pressure~\cite{Grest03}.   These values were used to estimate the
capillary number $Ca=\eta\,v_x/\gamma\approx 0.3$, which indicates
that the droplet motion is dominated by surface tension for all
values of the external force,
$f_x\leqslant0.00006\,\varepsilon/\sigma$, considered in the present
study.   For reference, the typical value of the Weber number is
$We=\rho\,v_x^2\,\ell/\gamma\approx0.1$, where $\ell \approx
180\,\sigma$ is the droplet size and $v_x=0.023\,\sigma/\tau$.



\vskip 0.05in

\begin{figure}[t]
\includegraphics[width=15.cm,angle=0]{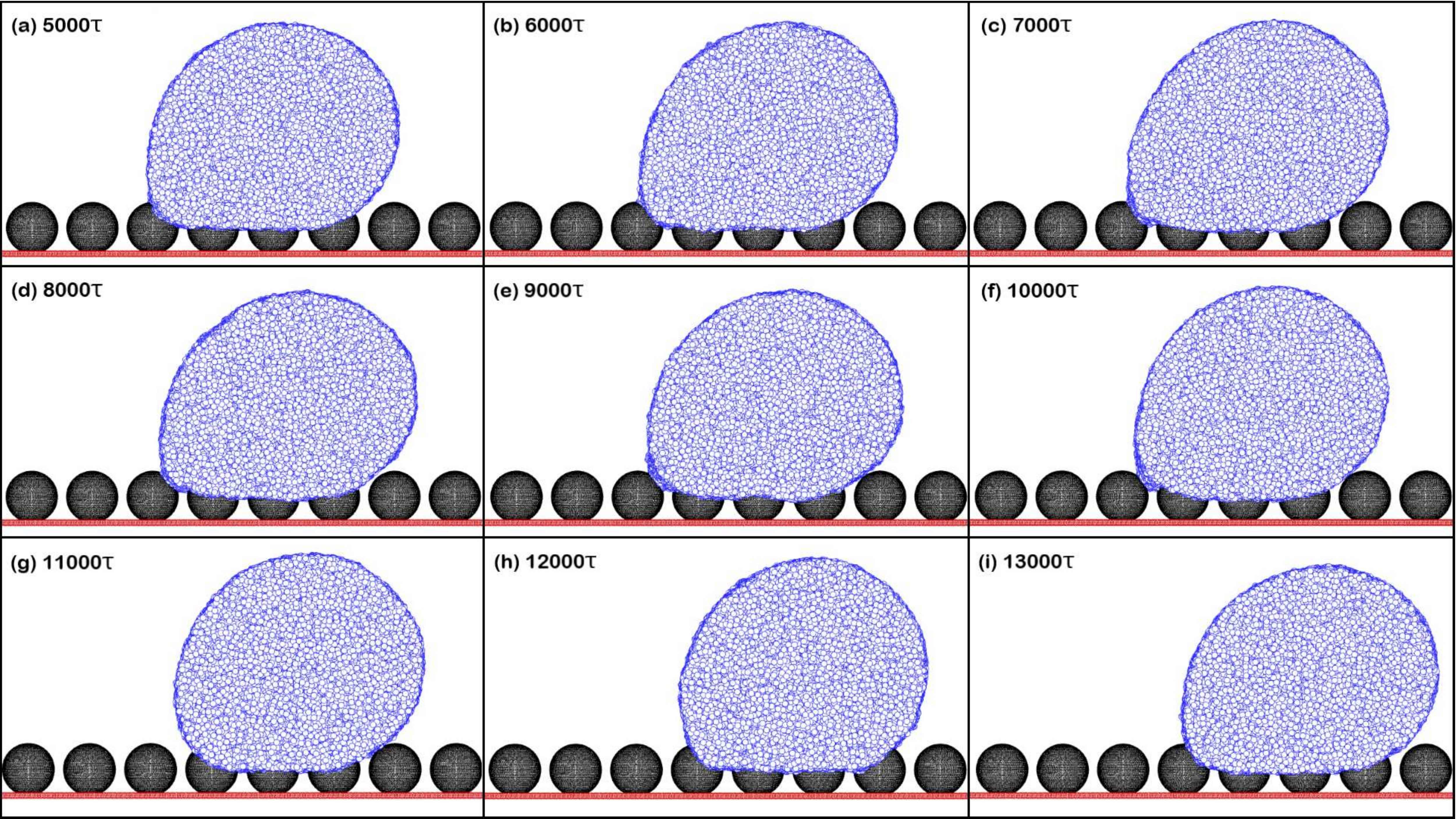}
\caption{(Color online) The sequence of snapshots illustrating the
advancement of the droplet front interface over the array of
particles. The external force is $f_x=0.00003\,\varepsilon/\sigma$
and the surface energy is $\varepsilon_{\rm
sp}\!=\!0.6\,\varepsilon$.  The time is measured after $f_x$ is
applied at $t=0$ (see the red curve in
Fig.\,\ref{fig:x_CM_time_fx}). }
\label{fig:adv_eps06_fx03}
\end{figure}
\begin{figure}[t]
\includegraphics[width=15.cm,angle=0]{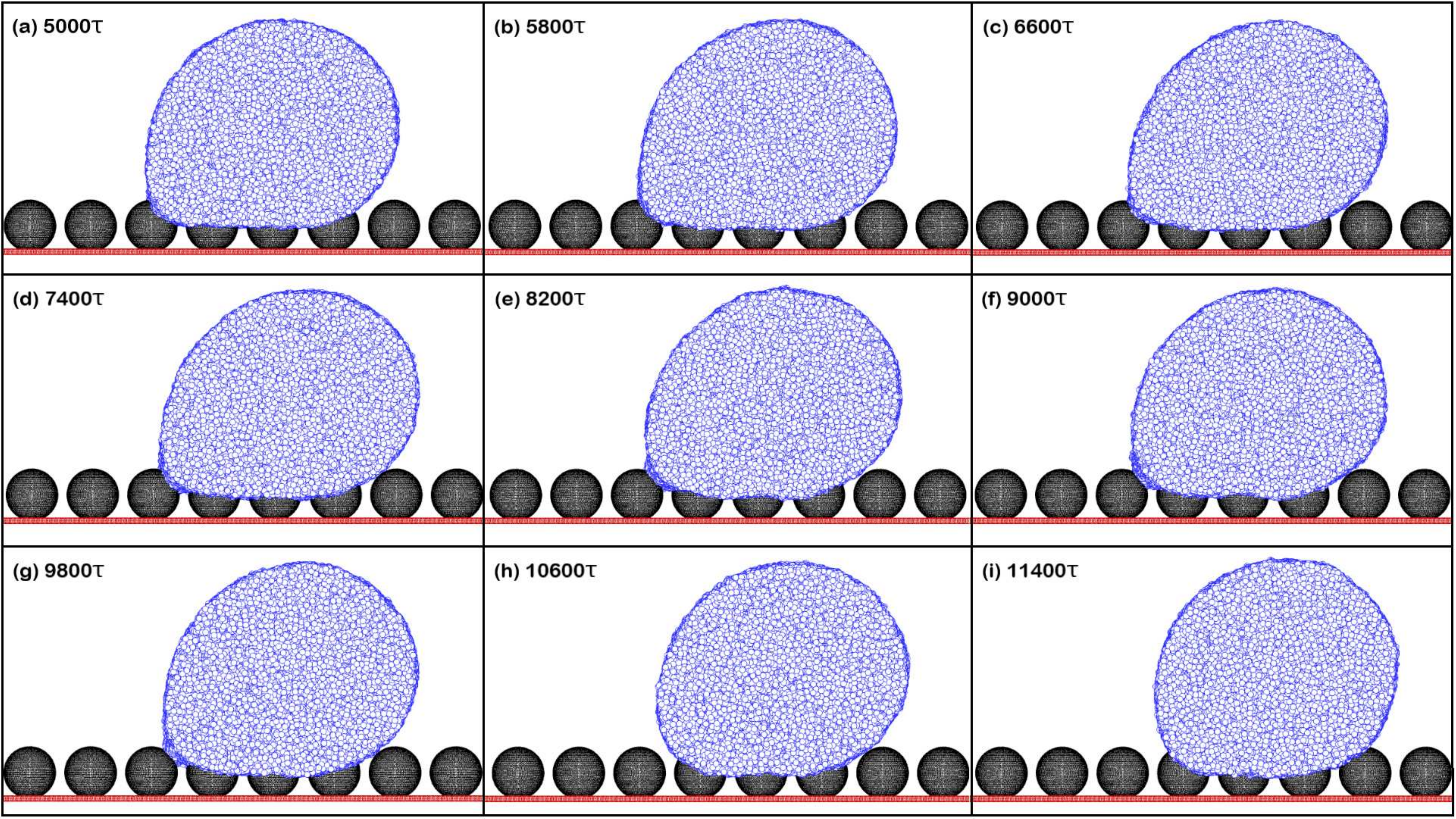}
\caption{(Color online) The process of detachment of the receding
contact line from the particle surface for the indicated time
intervals after the external force $f_x=0.00003\,\varepsilon/\sigma$
is applied on each monomer. The same data as in
Fig.\,\ref{fig:x_CM_time_fx} (the red curve) and in
Fig.\,\ref{fig:th_vx_time_eps06_fx03}.}
\label{fig:rec_eps06_fx03}
\end{figure}

The processes of attachment and detachment of the polymer interface
at the particles' surface are visualized in
Figs.\,\ref{fig:adv_eps06_fx03} and \ref{fig:rec_eps06_fx03}. The
snapshots of the droplet shape are taken at selected time intervals
after the external force $f_x=0.00003\,\varepsilon/\sigma$ is
applied at $t=0$. The displacement of the center of mass for these
cases is denoted by the red curve in Fig.\,\ref{fig:x_CM_time_fx}.
It can be observed in Fig.\,\ref{fig:adv_eps06_fx03}\,(b-g) that the
advancing interface remains effectively suspended in front of the
second rightmost particle during the extended time interval of about
$\Delta t = 5000\,\tau$, while the advancing contact angle is the
largest, $\theta_a\approx140^{\circ}$, when $t=9000\,\tau$ (see
Fig.\,\ref{fig:th_vx_time_eps06_fx03}).  Furthermore, the sequence
of snapshots in Fig.\,\ref{fig:rec_eps06_fx03}\,(a-g) shows the time
evolution of the circular contact line at the third particle which
temporarily restrains the droplet motion. Notice in
Fig.\,\ref{fig:rec_eps06_fx03}\,(e-g) that a narrow neck is formed
above the particle surface, followed by rapid detachment of the
polymer interface and displacement of the whole droplet along the
substrate.  These results suggest that one of the factors
determining the pinning time is the shape of surface texture; and,
thus, the droplet motion along the surface can be controlled by
introducing shape anisotropy of the surface-attached particles.


\vskip 0.05in

\begin{figure}[t]
\includegraphics[width=12.cm,angle=0]{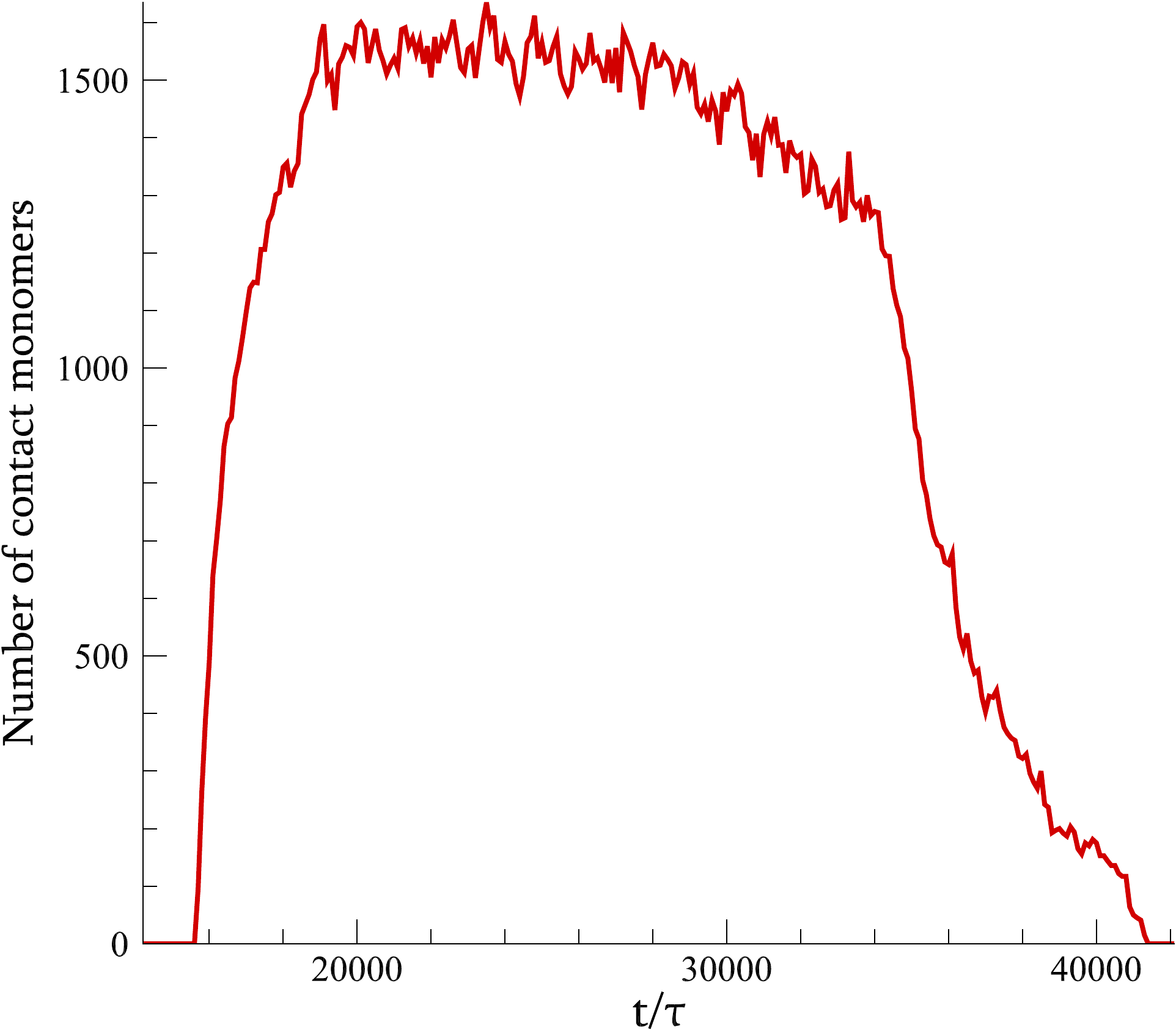}
\caption{(Color online) The time dependence of the number of fluid
monomers in contact with the surface of a spherical particle during
droplet motion under the external force
$f_x=0.00003\,\varepsilon/\sigma$. }
\label{fig:contact_atoms_fx03}
\end{figure}

A quantitative description of the attachment and detachment of the
droplet interface on the surface of a spherical particle can be
obtained by computing the number of fluid monomers in contact with
surface atoms.  More specifically, at any instant of time, we
identified fluid monomers that are located within the distance
$1.5\,\sigma$ from the position of an atom of a particular sphere.
The results are shown in Fig.\,\ref{fig:contact_atoms_fx03} for the
case $\varepsilon_{\rm sp}\!=\!0.6\,\varepsilon$ and
$f_x=0.00003\,\varepsilon/\sigma$.  As is evident, the number of
monomers in contact with one sphere first increases rapidly, then
saturates to a quasi-plateau, followed by a slow decay.  The sharp
increase in the number of contact monomers corresponds to a rapid
advancement of the droplet interface during step-like motion of the
center of mass (see the red curve in Fig.\,\ref{fig:x_CM_time_fx}).
In turn, the plateau level represents about half of the surface area
in contact with the liquid during droplet sliding along the
substrate.  Finally, the relatively slow decrease in the number of
contact monomers describes the detachment of progressively narrowing
neck that temporarily pins the whole droplet.

\section{Conclusions}

In summary, the dynamics of a polymer droplet driven over a
nanostructured interface composed of an array of spherical particles
was investigated using large-scale molecular dynamics simulations.
The areal density of nanoparticles and the interaction energy
between solid atoms and liquid monomers was adjusted to form a
partially wetting droplet with the apparent contact angle greater
than $90^{\circ}$ at mechanical equilibrium.  It was shown that the
polymer droplet remains pinned at the surface of nanoparticles at
sufficiently small values of the external force.  Above the
threshold force, the droplet moves intermittently along the
interface via successive rapid displacements and periods of pinning.
The process of attachment and detachment at the particles' surface
causes significant distortion of the droplet shape quantified via
advancing and receding contact angles.  With increasing external
force, the variation of the advancing contact angle is reduced,
whereas the oscillation amplitude of the receding contact angle and
the maximum contact angle hysteresis remain nearly unchanged.

\section*{Acknowledgments}


Financial support from the ACS Petroleum Research Fund (60092-ND9)
and the National Science Foundation (CNS-1531923) is gratefully
acknowledged. The molecular dynamics simulations were performed
using the LAMMPS open-source code developed at Sandia National
Laboratories~\cite{Lammps}.  The simulations were carried out at
Wright State University's Computing Facility and the Ohio
Supercomputer Center.



%

%

%

%

%

%

%

\bibliographystyle{prsty}

\begin{thebibliography}{99}

\bibitem{Sam19}       E.~K. Sam, D.~K. Sam, X. Lv, B. Liu, X. Xiao, S. Gong, W. Yu, J. Chen, and J. Liu,
                      Recent development in the fabrication of self-healing superhydrophobic surfaces,
                      Chem. Eng. {\bf 373}, 531 (2019).

\bibitem{Jeevahan18}  J. Jeevahan, M. Chandrasekaran, G.~B. Joseph, R.~B. Durairaj, and G. Mageshwaran,
                      Superhydrophobic surfaces: a review on fundamentals, applications, and challenges,
                      J. Coat. Technol. Res. {\bf 15}, 231 (2018).

\bibitem{Phuong19}    P. Nguyen-Tri, H.~N. Tran, C.~O. Plamondon, L. Tuduri, D.-V.~N. Vo, S. Nanda,
                      A. Mishra, H.-P. Chao, and A.~K. Bajpai,
                      Recent progress in the preparation, properties and applications of superhydrophobic
                      nano-based coatings and surfaces: A review,
                      Progress in Organic Coatings {\bf 132}, 235 (2019).


\bibitem{CB44}        A.~B.~D. Cassie and S. Baxter,
                      Wettability of porous surfaces,
                      Trans. Faraday Soc. {\bf 40}, 546 (1944).

\bibitem{Wenzel36}    R.~N. Wenzel,
                      Resistance of solid surfaces to wetting by water,
                      Ind. Eng. Chem. {\bf 28}, 988 (1936).

\bibitem{Vinograd12}  O.~I. Vinogradova and A.~L. Dubov,
                      Superhydrophobic textures for microfluidics,
                      Mendeleev Commun. {\bf 22}, 229 (2012).

\bibitem{Neto14}      T. Lee, E. Charrault, and C. Neto,
                      Interfacial slip on rough, patterned and soft surfaces:
                      A review of experiments and simulations,
                      Adv. Colloid Interface Sci. {\bf 21}, 210 (2014).



\bibitem{Priezjev05}  N.~V. Priezjev, A.~A. Darhuber, and S.~M. Troian,
                      Slip behavior in liquid films on surfaces of patterned wettability:
                      Comparison between continuum and molecular dynamics simulations,
                      Phys. Rev. E {\bf 71}, 041608 (2005).


\bibitem{Priezjev11}  N.~V. Priezjev,
                      Molecular diffusion and slip boundary conditions at smooth surfaces
                      with periodic and random nanoscale textures,
                      J. Chem. Phys. {\bf 135}, 204704 (2011).

%
\bibitem{Bao18suphyd}  H. Hu, D. Wang, F. Ren, L. Bao, N.~V. Priezjev, J. Wen,
                       A comparative analysis of the effective and local slip lengths for liquid flows over a trapped nanobubble,
                       Int. J. Multiph. Flow {\bf 104}, 166 (2018).

%
\bibitem{Bao20SH}      L. Bao, N.~V. Priezjev, H. Hu,
                       The local slip length and flow fields over nanostructured superhydrophobic surfaces,
                       Int. J. Multiph. Flow {\bf 126}, 103258 (2020).




\bibitem{Servantie08} J. Servantie and M. Muller,
                      Statics and dynamics of a cylindrical droplet under an external body force,
                      J. Chem. Phys. {\bf 128}, 014709 (2008)


\bibitem{Koishi09}    T. Koishi, K. Yasuoka, S. Fujikawa, T. Ebisuzaki, and X.~C. Zeng,
                      Coexistence and transition between Cassie and Wenzel state on
                      pillared hydrophobic surface,
                      Proc. Natl. Acad. Sci. USA {\bf 106}, 8435 (2009).


\bibitem{Yong09}      X. Yong and L.~T. Zhang,
                      Nanoscale wetting on groove-patterned surfaces,
                      Langmuir {\bf 25}, 5045 (2009).


\bibitem{ZhaoSci13}   Q. Yuan and Y.-P. Zhao,
                      Wetting on flexible hydrophilic pillar-arrays,
                      Sci. Rep. {\bf 3}, 1944, (2013).


\bibitem{Singh14}     S. Khan and J.~K. Singh,
                      Wetting transition of nanodroplets of water on textured surfaces: a molecular dynamics study,
                      Molecular Simulation {\bf 40}, 458 (2014).


\bibitem{Gendt14}     X.~M. Xu, G. Vereecke, C. Chen, G. Pourtois, S. Armini, N. Verellen,
                      W.-K. Tsai, D.-W. Kim, E. Lee, C.-Y. Lin, P.~V. Dorpe, H. Struyf,
                      F. Holsteyns, V. Moshchalkov, J. Indekeu, and S. De Gendt,
                      Capturing wetting states in nanopatterned silicon,
                      ACS Nano {\bf 8}, 885 (2014).

\bibitem{Tang14}      D. Niu and G.~H. Tang,
                      Static and dynamic behavior of water droplet on solid surfaces with
                      pillar-type nanostructures from molecular dynamics simulation,
                      Int. J. Heat Mass Transf. {\bf 79}, 647 (2014).

\bibitem{Pei15}       H.-W. Pei, H. Liu, Z.-Y. Lu, and Y.-L. Zhu,
                      Tuning surface wettability by designing hairy structures,
                      Phys. Rev. E {\bf 91}, 020401(R) (2015).


\bibitem{ZhaoNano15}  Q. Yuan and Y.-P. Zhao,
                      Statics and dynamics of electrowetting on pillar-arrayed surfaces at the nanoscale,
                      Nanoscale {\bf 7}, 2561 (2015).

\bibitem{Wang15}      J. Wang, S. Chen, and D. Chen,
                      Spontaneous transition of a water droplet from the Wenzel state to the Cassie state:
                      a molecular dynamics simulation study,
                      Phys. Chem. Chem. Phys. {\bf 17}, 30533 (2015).

\bibitem{Das16}       A.~M. Miqdad, S. Datta, A.~K. Das, and P.~K. Das,
                      Effect of electrostatic incitation on the wetting mode of
                      a nano-drop over a pillar-arrayed surface,
                      RSC Adv. {\bf 6}, 110127 (2016).


\bibitem{Ma16}        W. Xu, Z. Lan, B.~L. Peng, R.~F. Wen, and X.~H. Ma,
                      Effect of nano structures on the nucleus wetting modes during water
                      vapour condensation: from individual groove to nano-array surface,
                      RSC Adv. {\bf 6}, 7923 (2016).

\bibitem{Tsao16}      C.-C. Chang, Y.-J. Sheng, and H.-K. Tsao,
                      Wetting hysteresis of nanodrops on nanorough surfaces,
                      Phys. Rev. E {\bf 94}, 042807 (2016).

\bibitem{JYan17}      J. Yan, K. Yang, X. Zhang, and J. Zhao,
                      Analysis of impact phenomenon on superhydrophobic surfaces based on
                      molecular dynamics simulation,
                      Comput. Mater. Sci. {\bf 134}, 8 (2017).

\bibitem{Yen17}       T.-H. Yen,
                      Investigating the effects of wettability and gaseous nanobubbles on roughened
                      wall-fluid interface using molecular dynamics simulation,
                      Molecular Simulation {\bf 43}, 1 (2017).


\bibitem{Yu18}        H. Li, T. Yan, K.~A. Fichthorn, and S. Yu,
                      Dynamic contact angles and mechanisms of motion of water droplets moving on
                      nano-pillared superhydrophobic surfaces: A molecular dynamics simulation study,
                      Langmuir {\bf 34}, 9917 (2018).

\bibitem{Hendy18}     A.~F.~W. Smith, K. Mahelona, and S.~C. Hendy,
                      Rolling and slipping of droplets on superhydrophobic surfaces,
                      Phys. Rev. E {\bf 98}, 033113 (2018).

\bibitem{Prie20Geng}  X. Geng, X. Yu, L. Bao, N.~V. Priezjev, Y. Lu,
                      Directed transport of liquid droplets on vibrating substrates with asymmetric
                      corrugations and patterned wettability: A dissipative particle dynamics study,
                      Molecular Simulation {\bf 46}, 33 (2020).





\bibitem{Allen87}     M.~P. Allen and D.~J. Tildesley,
                      {\it Computer Simulation of Liquids} (Clarendon, Oxford, 1987).

\bibitem{Lammps}      S.~J. Plimpton,
                      Fast parallel algorithms for short--range molecular dynamics,
                      J. Comp. Phys. {\bf 117}, 1 (1995).

\bibitem{Kremer90}    K. Kremer and G.~S. Grest,
                      Dynamics of entangled linear polymer melts: A molecular dynamics simulation,
                      J. Chem. Phys. {\bf 92}, 5057 (1990).


\bibitem{McKinley07}  A. Tuteja, W. Choi, M. Ma, J.~M. Mabry, S.~A. Mazzella,
                      G.~C. Rutledge, G.~H. McKinley, and R.~E. Cohen,
                      Designing superoleophobic surfaces,
                      Science {\bf 318}, 1618 (2007).

\bibitem{Bishal18}    B. Bhattarai and N.~V. Priezjev,
                      Wetting properties of structured interfaces composed of surface-attached spherical nanoparticles,
                      Comput. Mater. Sci. {\bf 143}, 497 (2018).

\bibitem{Grest03}     D.~R. Heine, G.~S. Grest, and E.~B. Webb,
                      Spreading dynamics of polymer nanodroplets,
                      Phys. Rev. E {\bf 68}, 061603 (2003).



\end{thebibliography}

\end{document}